\documentclass[
aps,
prb,
preprint,
floatfix
]{revtex4-1}
\usepackage{graphicx}
\usepackage{multirow}

\begin{document}
\title{Structural stability and energy-gap modulation through atomic protrusion in freestanding bilayer silicene}
\author{Yuki Sakai}
\author{Atsushi Oshiyama}
\affiliation{Department of Applied Physics, The University of Tokyo, 7-3-1, Hongo, Bunkyo-ku, Tokyo 113-8656, Japan}
\date{\today}

\begin{abstract}
We report on first-principles total-energy and phonon calculations that clarify structural stability and electronic 
properties of freestanding bilayer silicene. By extensive structural exploration, we reach all the stable 
structures reported before and find four new dynamically stable structures, including the structure with the largest 
cohesive energy. We find that atomic protrusion from the layer is the principal relaxation pattern which stabilizes 
bilayer silicene and determines the lateral periodicity. The hybrid-functional calculation shows that the most 
stable bilayer silicene is a semiconductor with the energy gap of 1.3~eV. 
\end{abstract}


\maketitle

Bilayer graphene provides new aspects of graphene physics such as the band gap opening\cite{Ohta2006,McCann2006,
Castro2007,Zhang2009} and moir\'e-pattern-induced electron localization.\cite{Lucian2011,Uchida2014} 
The interlayer interaction is obviously weaker than the intralayer interaction but decisive to modulate the 
electron states due to its symmetry breaking. Similar intriguing behavior with the extension related to the 
spin-degrees of freedom is expected\cite{LiuPRL2011,Ezawa2012,Ezawa2013} for layered Si (silicene), which has been 
grown experimentally with a form of monolayer\cite{Vogt2012,Lin2012,Feng2012,Chiappe2012,Fleurence2012,
Jamgotchian2012,Guo2013-1,Guo2013-2} and of a few layers\cite{Arafune2013,Padova2013,Resta2013,Guo2014,Guo2015} 
mainly on Ag substrates. One of the most important characteristics which discriminate silicene from graphene is 
the buckling of two sublattices caused by the preference of Si for $sp^3$ hybridization. In fact, first-principles 
calculations within the local density approximation\cite{Takeda1994,Cahangirov2009} have clarified 
that a planar Si monolayer is unstable to the buckling, and that the resultant 
freestanding monolayer silicene with the buckling of 0.44~{\AA} have the Dirac cone at the Fermi 
level, $E_{\rm F}$. This buckling brings about a complex but rich variation in structure and 
electronic properties in bilayer silicene, that we address in this Rapid Communication. 

Several theoretical investigations have been performed on freestanding bilayer 
silicene.\cite{Bai2010,Morishita2011,Fu2014,Lian2013,Luo2014,Huang2014,Padilha2015}
A flat bilayer structure with the perfect overlapping stacking ($AA$ stacking) 
has been predicted based on a molecular dynamics simulation.\cite{Bai2010}
A corrugated $2\times2$ (rectangular supercell) reconstructed structure 
different from the well known $\pi$-bonded chain structure of the Si(111) surface\cite{Pandey1981} 
has been also found.\cite{Morishita2011} Recently, freestanding bilayer silicene 
has been found to have several local minima in total energy as a function of the lateral lattice parameter,
suggesting its wealth of the structural diversity.\cite{Fu2014} 
A possible structural phase transition under the lateral strain has been also discussed.\cite{Lian2013} 
To validate such theoretical predictions, however, a systematic exploration of stable 
structures with their thermal excitation spectra (phonons), which is lacking in the past, is imperative. 
Further, clarification of correlation between the structural 
diversity and its role in electronic properties is highly demanded. 

We here perform systematic first-principles total-energy and phonon calculations for freestanding bilayer 
silicene with various lateral periodicities and atomic densities. By extensive geometry optimization followed 
by the phonon calculations, we unequivocally identify ten dynamically stable structures with distinct atomic 
configurations, symmetries, and periodicities. 
We clarify that the ten structures include all the six structures reported in the past. 
Other four structures are newly found and more stable than the previously reported ones. 
We find that the more stable structures have a single prominent structural characteristics, i.e. the protrusion
of Si atom. We also find that there is an energetically optimum lateral periodicity for the protruded structure, 
i.e., $\sqrt{3}\times\sqrt{3}$ or $2\times2$, depending on the stacking of the two Si layers. 
We further clarify that the stable freestanding bilayer silicene is a semiconductor in which the energy gap is 
sensitive to the detailed protruded structure. 

We use the pseudopotential-planewave method\cite{Troullier1991, Giannozzi2009} based on the 
density-functional theory.\cite{Hohenberg1964, Kohn1965} We adopt the exchange-correlation functionals developed 
by Perdew, Burke, and Erenzerhof (PBE).\cite{Perdew1996} We crosscheck the electronic properties by using the 
Heyd-Scuseria-Ernzerhof (HSE) hybrid functional.\cite{Heyd2003, Heyd2006} Phonon dispersion relations are calculated 
based on the density functional perturbation theory.\cite{Baroni1987, Gonze1995, Baroni2001}
Computational details are described in Supplemental Material.\cite{SM}

In our calculations, we have considered $1\times1$, $2\times2$, $\sqrt{3}\times\sqrt{3}$ and $\sqrt{7}\times\sqrt{7}$ 
lateral periodicities with respect to the monolayer periodicity and performed total-energy optimization. 
Freestanding monolayer silicene is intrinsically buckled so that the stacking of two Si monolayers is of rich variety: 
the buckling of two layers may be in-phase, out-of-phase, or planar both in the $AA$ stacking and in the $AB$ stacking 
(Bernal stacking named in graphite); further the stacking is not restricted to $AA$ or $AB$ but could be orthorhombic 
(OR) in general. We have then noticed that there are 16 distinct stacking configurations.\cite{SM} 
For each stacking configuration, we have adopted a certain lateral lattice parameter and fully optimized atomic 
geometries, and then repeated the calculations with varying the lattice parameter. We have also examined a possibility 
of the unit-cell distortion. By such extensive exploration, we have obtained 24 distinct total-energy minimized geometries 
for the $1\times1$, the $2\times2$ and $\sqrt{3}\times\sqrt{3}$ periodicities. We have then calculated the phonon spectra 
for thus obtained total-energy minimized structures. Surprisingly, the 14 of the 24 total-energy minimized structures is 
clarified to be unstable with imaginary phonon frequencies, although their interlayer binding energies are positive. 
We have then eventually reached four $1\times1$, four $2\times2$ and two $\sqrt{3}\times\sqrt{3}$ dynamically 
stable structures. Details of our structural exploration are described in Supplemental Material.\cite{SM}

\begin{figure}[htp]
  \includegraphics[width = 8.6cm]{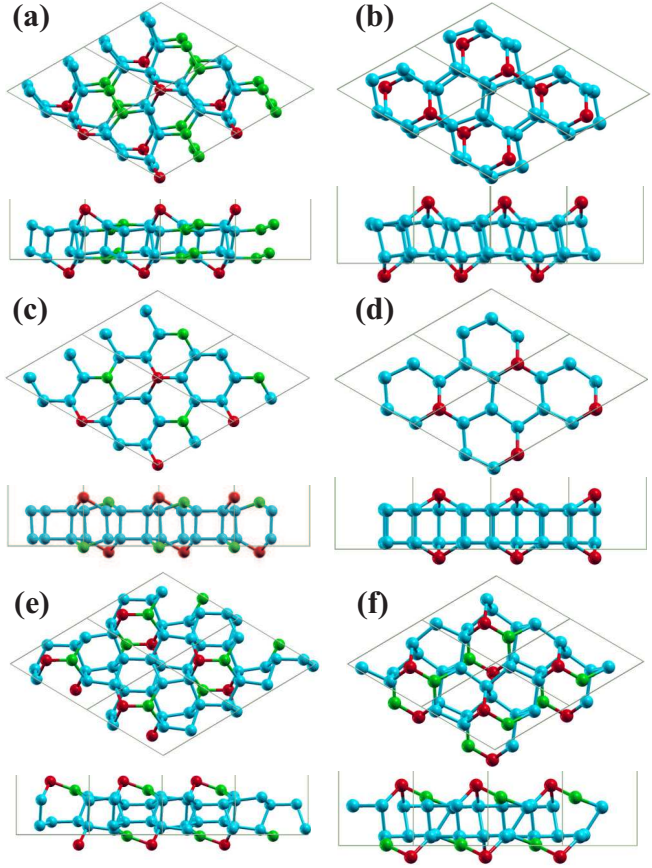}
    \caption{
      \label{fig:structure}
(Color online)
Total-energy minimized structures of $\sqrt{3}\times\sqrt{3}$ and 2$\times$2 freestanding bilayer silicene: (a) slide-2$\times$2, 
(b) slide-$\sqrt{3}\times\sqrt{3}$, (c) $AA$-2$\times$2, (d) $AA$-$\sqrt{3}\times\sqrt{3}$, (e) hex-OR-2$\times$2, 
and (f) OR-$\sqrt{3}\times\sqrt{3}$ structures. Unit cells are represented by silver lines.
Blue, red, and green spheres represent fourfold coordinated, highly-protruded, and moderately-protruded silicon atoms, respectively.
The angle between two lattice vectors are slightly changed from 120$^\circ$ in the slide-2$\times$2, 
slide-$\sqrt{3}\times\sqrt{3}$, hex-OR-2$\times$2, and OR-$\sqrt{3}\times\sqrt{3}$ cases.
The XCrySDen program is used for visualization of the atomic structures.\cite{Kokalj2003}}
\end{figure}

The five dynamically stable structures obtained in the present study shown in Fig.~S4 (Supplemental Material\cite{SM}) and 
Fig.~\ref{fig:structure}(a), which are named $AA$-1$\times$1, $AB$-1$\times$1, slide-1$\times$1, OR-1$\times$1 and 
slide-2$\times$2, are presumably identical to the structures which have been reported 
previously.\cite{Morishita2011, Lian2013, Fu2014, Luo2014, Huang2014} The $AA$-1$\times$1 structure is similar to the 
$AA$-stacking bilayer graphene, but the interlayer distance is significantly small (2.411~{\AA}), resulting in the 
fourfold coordination of all the Si atoms. The $AB$-1$\times$1 structure is similar to an atomic slab in the diamond 
structure, inferring the existence of threefold coordinated Si. The slide-1$\times$1 structure also shows threefold 
and fourfold coordinated atoms, although the interlayer distance becomes shorter compared to the $AB$-1$\times$1. 
The OR-1$\times$1 structure is obtained by relaxing the lateral and in-plane bond angles of $AA$-1$\times$1 structure.
The $AA$-1$\times$1, $AB$-1$\times$1, 
slide-1$\times$1, OR-1$\times$1, and slide-2$\times$2 structures are presumably identical to the 1AA-, the 1AB-, 
the slide-2AA-structures in ref.~\onlinecite{Fu2014}, the Si-Cmme structure in ref.~\onlinecite{Luo2014}, and the phase 
II structure obtained in ref.~\onlinecite{Lian2013}, respectively. We have also obtained the structure proposed 
in ref.~\onlinecite{Morishita2011}, here denoted by rect-OR-2$\times$2 [Fig.~S4(e)], by using a rectangular $2\times2$ supercell. 
The slide-2$\times$2 and the rect-OR-2$\times$2 structures exhibit peculiar structural characteristics: 
two Si atoms in each unit cell protrudes prominently, causing the local tetrahedral geometry. 
It is noteworthy that this protrusion renders those two structures lower in energy than other 4 structures (see below). 

\begin{figure}[htp]
  \includegraphics[width = 8.6cm]{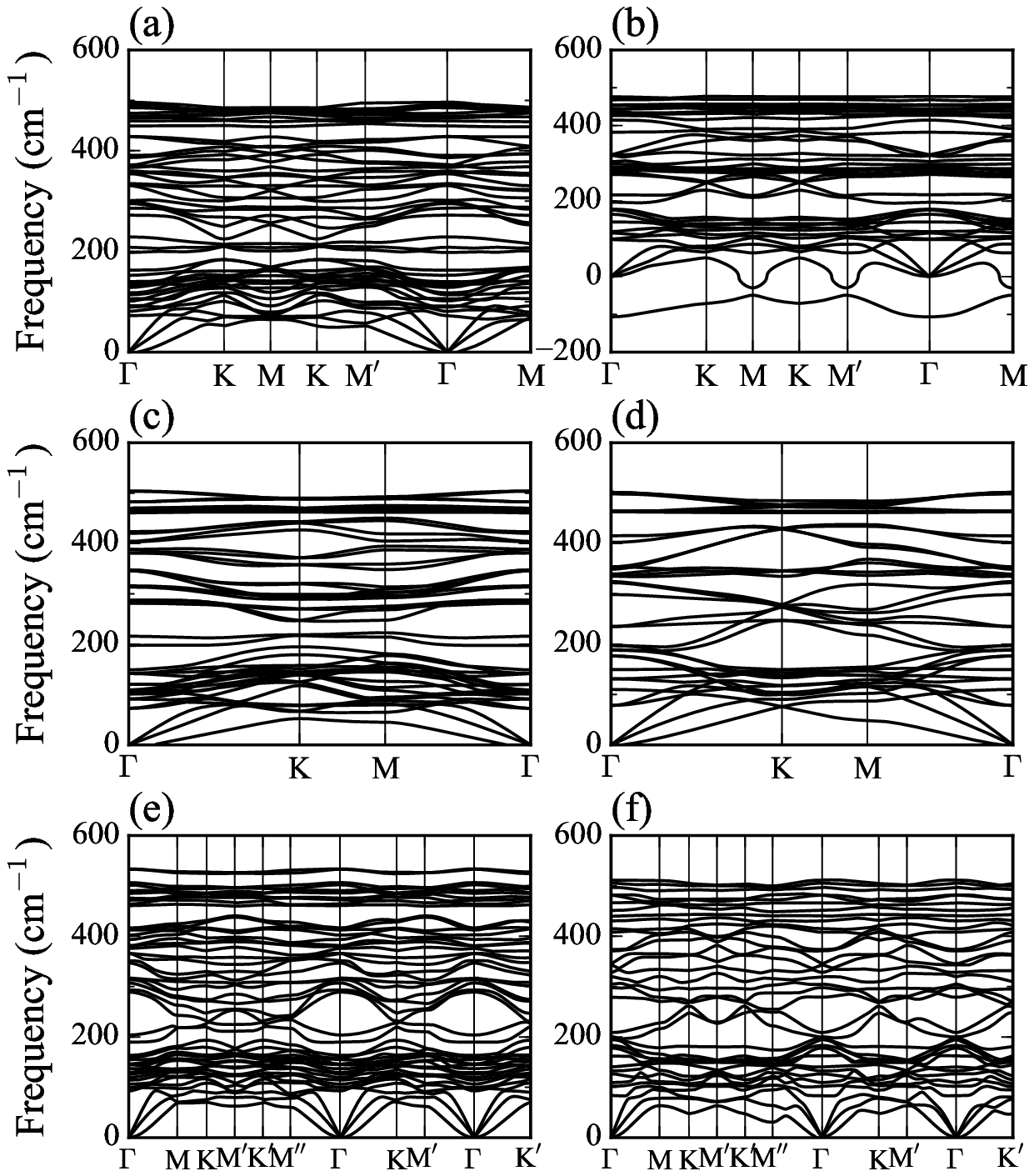}
    \caption{
    \label{fig:phonon}
Phonon dispersion relations of (a) slide-2$\times$2, (b) slide-$\sqrt{3}\times\sqrt{3}$, (c) $AA$-2$\times$2, (d) 
$AA$-$\sqrt{3}\times\sqrt{3}$, (e) hex-OR-2$\times$2, and (f) OR-$\sqrt{3}\times\sqrt{3}$ structures (in the same order 
as Fig.~\ref{fig:structure}). Phonon dispersions crossing several $M$ points, $M(0.5, 0)$, $M^{\prime}(0, 0.5)$ and 
$M^{\prime\prime}(-0.5, 0.5)$ in units of reciprocal lattice vector, which are inequivalent to each other in the low 
symmetry cases are plotted. The negative values in (b) mean imaginary phonon frequencies, 
indicating that the structure is dynamically unstable. Imaginary frequency branches (acoustic vibrational modes along the direction 
perpendicular to the layers) appear around the $\Gamma$ point in other cases due to the interpolation errors.  
We have confirmed that these imaginary frequencies become real when we directly compute the phonon frequency at specific 
$\mathbf{q}$ points. with sufficient vacuum space (16~{\AA}) of the supercell and cutoff energy (60~Ry).}
\end{figure}

In the 2$\times$2 periodicity, we have found two new structures, named $AA$-2$\times$2 [Fig.~\ref{fig:structure}(c)] 
and hex-OR-2$\times$2 [Fig.~\ref{fig:structure}(e)]. The latter is the lowest in energy among the structures 
ever reported (see below). The $AA$-2$\times$2 structure shows common characteristics to the slide-2$\times$2 structure 
[Fig.~\ref{fig:structure}(a) ], but the details are different. The $AA$-2$\times$2 structure possesses almost complete 
$AA$ stacking whereas the slight dislodgment is observed in the slide-2$\times$2. In the $AA$-2$\times$2, two silicon 
atoms in each unit cell are highly protruded [red spheres in Fig.~\ref{fig:structure}(c)] whereas another two atoms 
are moderately protruded (green spheres). In the hex-OR-2$\times$2 structure, on the other hand, 
red and green protruded Si atoms in Fig.~\ref{fig:structure}(e) form dimers.
The hex-OR-2$\times$2 
structure is similar to the rect-OR-2$\times$2 structure although the supercell is not rectangular but hexagonal. 
The four 2$\times$2 structures, including newly found two structures, are dynamically stable, as is evidenced from their 
calculated phonon spectra shown in Figs.~\ref{fig:phonon}(a), \ref{fig:phonon}(c) and \ref{fig:phonon}(e), and Fig.~S4(j). 

We have further explored dynamically stable structures with different lateral periodicity. Focusing the dynamically stable 
stacking configurations in the 2$\times$2 periodicity shown in Figs.~\ref{fig:structure} (a), (c) and (e), we have 
examined the $\sqrt{3}\times\sqrt{3}$ lateral periodicity with those stackings. We have found that the $AA$- and 
OR-$\sqrt{3}\times\sqrt{3}$ structures [Figs.~\ref{fig:structure}(d) and \ref{fig:structure}(f)] are also dynamically 
stable whereas the slide-$\sqrt{3}\times\sqrt{3}$ structure [Fig.~\ref{fig:structure}(b)] is unstable, as is evidenced in 
our phonon dispersion relations in Figs.~\ref{fig:phonon}(b), \ref{fig:phonon}(d) and \ref{fig:phonon}(f). 
To our knowledge, the $AA$- and the OR-$\sqrt{3}\times\sqrt{3}$ structures found here have not been reported in the past. 

The calculated total energies of freestanding bilayer silicene with various stackings and periodicities are listed 
in Table~\ref{tb:energy}. The hex-OR-2$\times$2 structure newly found here has the lowest energy, and the slide-2$\times$2 
follows with the total-energy increase of 6~meV/atom. The OR-$\sqrt{3}\times\sqrt{3}$ and rect-OR-2$\times$2 structures 
are also close in energy (+9 and +10~meV/atom from the lowest energy, respectively). The $AA$-$\sqrt{3}\times\sqrt{3}$ 
and $AA$-2$\times$2 are in the second lowest energy group (+31 and +41~meV/atom, respectively). 
The OR-, $AA$-, slide-, $AB$-1$\times$1 structures have relatively high energy
(+60, +78, +112 and +131~meV/atom, respectively) compared to the $\sqrt{3}\times\sqrt{3}$ and 2$\times$2 structures.
We have also calculated the total energy of another structure named 
honeycomb dumbbell silicene,\cite{Cahangirov2014} constructed from a $3\times3$ monolayer silicene. 
This dumbbell structure is outside our configurational phase space
for the structural search, although it has higher total energy than the value of the hex-OR-2$\times$2 structure 
by 206~meV/atom ($\sqrt{3}\times\sqrt{3}$ periodicity) or 60~meV/atom (2$\times$2 periodicity). 
\begin{table}[htp]
\caption{\label{tb:energy}
Calculated total energies (in meV/atom) of various stacking geometries ($AA$, slide, OR, and $AB$) and lateral 
periodicities (1$\times$1, $\sqrt{3}\times\sqrt{3}$, 2$\times$2, and $\sqrt{7}\times\sqrt{7}$) of freestanding bilayer 
silicene. The total energies are compared to that of freestanding monolayer silicene. We do not find a 
distinct $AB$-stacking structure for the $\sqrt{3}\times\sqrt{3}$ and 2$\times$2 periodicities. 
}
\begin{ruledtabular}
\begin{tabular}{ccccc}
	\multirow{2}{*}{Stacking} &         \multicolumn{4}{c}{Periodicity} \\ \cline{2-5}
	                      & 1$\times$1 & $\sqrt{3}\times\sqrt{3}$ & 2$\times$2 & $\sqrt{7}\times\sqrt{7}$\\
\hline
$AA$      & $-183$       & $-230$                 & $-220$     & $-217$                  \\
Slide     & $-149$       & $-234$                 & $-255$     & $-223$                  \\
\multirow{2}{*}{OR} & \multirow{2}{*}{$-201$} & \multirow{2}{*}{$-252$} & $-261$ (hex) & \multirow{2}{*}{$-219$} \\
                    &                         &                         & $-251$ (rect)&  \\
$AB$      & $-130$       &  -                     & -          &  -                      \\
\end{tabular}
\end{ruledtabular}
\end{table}

Our systematic total-energy and phonon calculations have clarified rich structural variety of bilayer silicene. 
At the same time, it has become clear that stable bilayer silicene is accompanied with a characteristic relaxation, 
i.e., the atomic protrusion. The $\sqrt{3}\times\sqrt{3}$ structures and their structural analogues with the 
2$\times$2 periodicity have common stacking configurations and protruded structures but different total energies. 
This implies the existence of an optimum periodicity for the protruded structure for each stacking configuration. 
In order to clarify this point, we have performed total energy calculations for the $\sqrt{7}\times\sqrt{7}$ 
structures having the characteristic protruded pattern for the three stacking geometries ($AA$, slide and OR). 
The total energies of the fully optimized $\sqrt{7}\times\sqrt{7}$ structures are shown in Table~\ref{tb:energy}. 
They are all higher than the corresponding total energies of the 2$\times$2-periodicity cases. The total energies of 
the slide and OR stacking geometries become minimum with the $2\times$2 periodicity, whereas the total energy of 
the $AA$-stacking is minimum at the $\sqrt{3}\times\sqrt{3}$ periodicity. We have now unveiled that the atomic 
protrusion is the principal relaxation pattern which stabilizes the bilayer silicene and induces a particular 
lateral periodicity. 

\begin{figure}[htp]
  \includegraphics[width = 8.6cm]{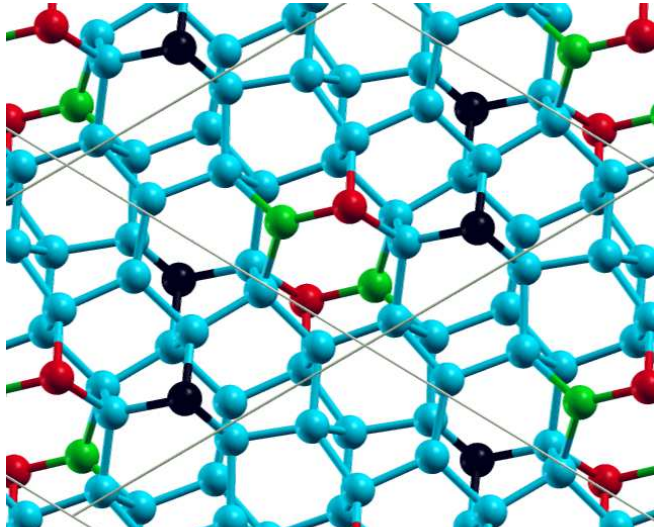}
    \caption{
    \label{fig:R7}
(Color online) Top view of OR-$\sqrt{7}\times\sqrt{7}$ structure. Red and green spheres are highly-protruded 
and moderately-protruded Si atoms as in Figs.~\ref{fig:structure}(e) and \ref{fig:structure}(f). Black spheres 
represent additionally protruded Si atoms.}
\end{figure}

Figure~\ref{fig:R7} shows the top view of the OR-$\sqrt{7}\times\sqrt{7}$ structure obtained here. Red and green 
spheres represent dimerized protrusion which is common to other periodicities. We now find that extra Si protrusions 
(black spheres) are induced in the $\sqrt{7}\times\sqrt{7}$ periodicity which are absent in the shorter periodicity. 
The protrusion generates the local tetrahedral geometry which is favorable for Si but at the same time induces 
stress energy around. The structure with the $\sqrt{7}\times\sqrt{7}$ periodicity indicates that the incomplete 
protrusion is energetically unfavorable. Similarly, the $AA$- and slide-$\sqrt{7}\times\sqrt{7}$ structures also show 
the extra protrusions. Such extra protrusions are also found in the $AA$-2$\times$2 structure where four of 16 Si 
atoms are protruded in each cell while the $AA$-$\sqrt{3}\times\sqrt{3}$ has one protruded atom in six Si atoms 
[Figs.~\ref{fig:structure}(c) and \ref{fig:structure}(d)]. This excess protrusion is also observed in the 
highest-energy $AB$-1$\times$1 structure where half of Si atoms are protruded. The hex-OR-2$\times$2 structure 
which we have found has the lowest energy is free from such extra protrusion. 

\begin{figure}[htp]
  \includegraphics[width = 8.6cm]{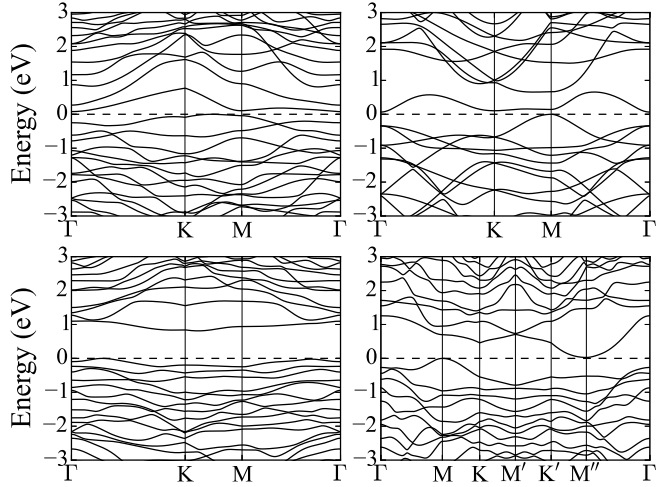}
    \caption{
    \label{fig:BS}
PBE band structures of dynamically stable (a) slide-2$\times$2, (b) $AA$-$\sqrt{3}\times\sqrt{3}$, (c) 
hex-OR-2$\times$2, and (d) OR-$\sqrt{3}\times\sqrt{3}$ structures. The horizontal dashed line indicates the 
valence band top. The path is chosen so that valence band top and conduction-band bottom can be seen in the plot.}
\end{figure}

Figure~\ref{fig:BS} shows the PBE band structures of the four most stable bilayer silicene, 
slide-2$\times$2, $AA$-$\sqrt{3}\times\sqrt{3}$, OR-$\sqrt{3}\times\sqrt{3}$ and hex-OR-2$\times$2. The total energy 
differences of these four structures are within the range of 31~meV/atom, indicating that they are stable at 
room temperature. The slide-2$\times$2, $AA$-$\sqrt{3}\times\sqrt{3}$, and OR-$\sqrt{3}\times\sqrt{3}$ structures 
are narrow gap semiconductors with indirect band gaps of 0.07 ($K$-$M$ line $\rightarrow \Gamma$), 0.06 
($M \rightarrow \Gamma$), and 0.02~eV ($M \rightarrow M^{\prime\prime}$), respectively. The band gaps of the 
three structures become approximately 0.37, 0.34 and 0.40~eV, respectively, when we use the hybrid HSE functional. 
Interestingly, the indirect band gap ($\Gamma$-$K$ line $\rightarrow$ $K$-$M$ line) of the most stable 
hex-OR-2$\times$2 structure is 0.8~eV (about 1.3~eV in the HSE calculation) considerably wide compared 
to other structures. 

\begin{figure}[htp]
  \includegraphics[width = 8.6cm]{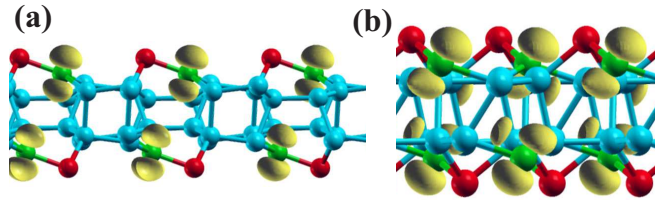}
    \caption{
    \label{fig:WFC}
  (Color online) Conduction band bottom Kohn-Sham wavefunctions of (a) hex-OR-2$\times$2 and (b) 
  OR-$\sqrt{3}\times\sqrt{3}$ structures. The yellow isosurfaces represent 1/3 of the maximum amplitude.}
\end{figure}

The hex-OR-2$\times$2 and OR-$\sqrt{3}\times\sqrt{3}$ structures have common protrusion but remarkably different 
band gaps. The conduction-band bottom state is more dispersive in the $\sqrt{3}\times\sqrt{3}$ structure compared 
to the 2$\times$2 case. In fact, the Kohn-Sham wavefunction at the conduction-band bottom state of the 
OR-$\sqrt{3}\times\sqrt{3}$ structure have amplitude at unprotruded Si atoms whereas the conduction-bottom state is 
highly-localized at the green protruded Si atoms in the 2$\times$2 case (Fig.~\ref{fig:WFC}). The deviation of the 
conduction-band bottom state can be related to the short distance between the protruded atoms in the 
$\sqrt{3}\times\sqrt{3}$ case. The band gap reduction is caused by the dispersive conduction band induced by 
the interaction between the protruded atoms. 

In summary, we have performed density-functional calculations to clarify atomic and electronic structures of 
freestanding bilayer silicene. After the extensive structural exploration, we have obtained four 1$\times$1, 
two $\sqrt{3}\times\sqrt{3}$, and four 2$\times$2 dynamically stable structures free from imaginary 
phonon frequencies. We have found that the atomic protrusion is a principal structural characteristics which 
stabilizes bilayer silicene and induces optimum lateral periodicity. We have identified the most stable bilayer 
silicene as the hex-OR-2$\times$2 structure and revealed that this is a semiconductor with the energy 
gap of 1.3~eV using the hybrid functional. 

\begin{acknowledgments}
We acknowledge Professor Z.-X.~Guo for helpful discussions.
This work was supported by the research project Materials Design through Computics 
(http://computics-material.jp/index-e.html) 
by MEXT and also by Computational Materials Science Initiative by MEXT, Japan.
Theoretical calculations were partly performed using Research Center for Computational 
Science, Okazaki, Japan.
\end{acknowledgments}
\end{document}